\documentclass[preprint,11.5pt]{elsarticle2}

\usepackage{amsmath}
\usepackage{mathrsfs}
\usepackage{amsfonts}
\usepackage{amssymb}
\usepackage{geometry}
 \usepackage[none]{hyphenat}
\usepackage{slashed}
\usepackage{amssymb}
\usepackage{amscd}
\usepackage{enumerate}
\usepackage{graphicx}
\usepackage{float}
\usepackage[lofdepth,lotdepth]{subfig}
\usepackage[countmax]{subfloat}
\usepackage{color}
\usepackage{etoolbox}
\usepackage{multirow}
\biboptions{sort&compress} % needed to change cites [1,2,...,N] to [1--N]

\usepackage{hyperref}
\definecolor{nicered}{rgb}{.7,.1,.1}
\definecolor{nicegreen}{rgb}{.1,.5,.1}
\definecolor{darkblue}{rgb}{0,0,.5}
\hypersetup{colorlinks, citecolor=nicegreen,linkcolor=nicered, urlcolor=darkblue}

\def\be{\begin{equation}}
\def\ee{\end{equation}}

\newcommand{\epem}{e^+e^-}

\newcommand{\mumu}{\mu^+\mu^-}
\newcommand{\tautau}{\tau^+\tau^-}
\newcommand{\gaga}{\gamma\gamma}

\newcommand{\sqrts}{\sqrt{s}}

\newcommand{\sqrtsnn}{\sqrt{s_{_\text{NN}}}}

\newcommand{\pT}{p_\text{T}}
\newcommand{\jpsi}{J/\psi}

\def\ttt#1{\texttt{\small #1}}

\newcommand{\Lumi}{\mathcal{L}}

\newcommand{\madgraph}{\textsc{MadGraph5\_aMC@NLO}}
\newcommand{\mgshort}{\textsc{MG5\_aMC}}
\newcommand{\gammaUPC}{\ttt{gamma-UPC}}

\newcommand{\superchic}{\textsc{Superchic}}

\newcommand{\ufo}{\textsc{ufo}}

\usepackage{xspace}
\newcommand*{\eg}{e.g.,\@\xspace}
 
\newcommand*{\cm}{c.m.\@\xspace}

\journal{XX}

\begin{document}

\begin{frontmatter}
\title{Collider constraints on massive gravitons coupling to photons}

\author{David d'Enterria$^{a,*}$, Malak Ait Tamlihat$^{b,\star}$, Laurent Schoeffel$^{c,\dagger}$, Hua-Sheng~Shao$^{d,\ddagger}$, Yahya~Tayalati$^{b,e,\mathsection}$}
\address{$a$ CERN, EP Department, CH-1211 Geneva, Switzerland\\
$b$ Mohammed V University in Rabat, Faculty of Sciences, 4 av. Ibn Battouta, B.P. 1014, R.P. 10000 Rabat, Morocco\\
$c$ Irfu, CEA, Université Paris-Saclay, 91191 Gif-sur-Yvette, France \\ 
$d$ Laboratoire de Physique Th\'eorique et Hautes Energies (LPTHE), UMR 7589,\\ Sorbonne Universit\'e et CNRS, 4 place Jussieu, 75252 Paris Cedex 05, France\\
$e$ Institute of Applied Physics, Mohammed VI Polytechnic University, Lot 660,\\ 43150 Hay Moulay Rachid Ben Guerir, Morocco
}
\vspace{20mm}
\address{$^*$david.d'enterria@cern.ch, $^\star$malak.ait.tamlihat@cern.ch, $^\dagger$laurent.olivier.schoeffel@cern.ch, $^\ddagger$huasheng.shao@lpthe.jussieu.fr, 
$^\mathsection$Yahya.Tayalati@cern.ch.
}

\begin{abstract}
\noindent
We study the discovery potential of massive graviton-like spin-2 particles coupled to standard model fields, produced in photon-photon collisions at the Large Hadron Collider (LHC) as well as in electron-positron ($\epem$) collisions, within an effective theory with and without universal couplings. Our focus is on a massive graviton G coupled to the electromagnetic field, 
which decays via $\mathrm{G}\to\gaga$ and leads to a resonant excess of diphotons over the light-by-light scattering continuum at the LHC, and of triphoton final states at $\epem$ colliders. Based on similar searches performed for pseudoscalar axion-like particles (ALPs), and taking into account the different cross sections, $\gaga$ partial widths, and decay kinematics of the pseudoscalar and tensor particles, we reinterpret existing experimental bounds on the ALP-$\gamma$ coupling into G-$\gamma$ ones. Using the available data, exclusion limits on the graviton-photon coupling are set down to $g_{\mathrm{G}\gamma}\approx 1$--0.05~TeV$^{-1}$ for masses $m_\mathrm{G} \approx 100$~MeV--2~TeV. 
Such bounds can be improved by factors of 100 at Belle~II in the low-mass region, and of 4 at the HL-LHC at high masses, with their expected full integrated luminosities.
\end{abstract}

%%\keywords{QED, Equivalent Photon Approximation, LHC}

\end{frontmatter}

%%%%%%%%%%%%%%%%%%%%%%%%%%%%%%%%%%
\section{Introduction}
%%%%%%%%%%%%%%%%%%%%%%%%%%%%%%%%%%

\begin{figure}[!htbp]
\centering
 \includegraphics[width=1.\textwidth]{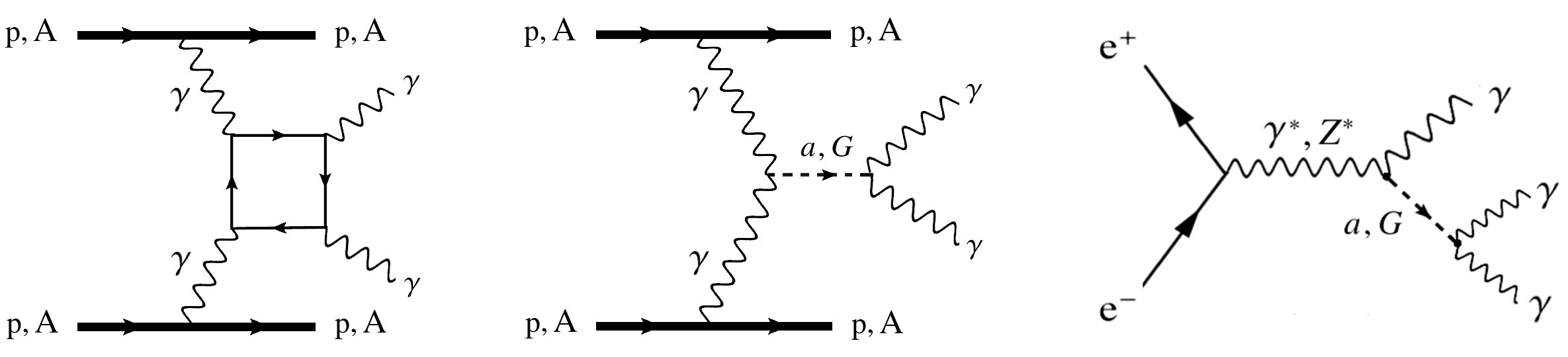}
 \caption{Schematic diagrams of photon-photon collisions producing a pair of exclusive photons, aka.\ LbL scattering (left), and an ALP or graviton decaying to two photons (center), and of $\epem$ collisions producing an ALP or graviton leading to a triphoton final state (right).}
 \label{fig:diags}
\end{figure}

The CERN Large Hadron Collider (LHC) does not only provide the highest energy and luminosity hadronic interactions recorded to date, but also delivers the most intense and energetic photon-photon collisions ever studied in the laboratory. 
In proton-proton, proton-ion, and ion-ion collisions, the ultrarelativistic beam charged particles can interact electromagnetically through photon exchange when passing by at large impact parameters (ultraperipheral collisions, UPCs) without hadronic overlap, and remain intact after the interaction~\cite{Baltz:2007kq,deFavereaudeJeneret:2009db}. In the equivalent photon approximation (EPA)~\cite{Brodsky:1971ud,Budnev:1975poe}, the collision of the two electromagnetic (EM) fields can be identified with the fusion of two quasireal photons, which can produce particles in the central detectors of the LHC experiments. Pairs of bosons or fermions can thus be produced, back-to-back in azimuth, via $\gamma\gamma$ processes (Fig.~\ref{fig:diags}, left) and ---by virtue of the Landau--Yang theorem~\cite{Landau:1948kw,Yang:1950rg} and conservation of charge-conjugation (C) symmetry--- C-even neutral objects (scalars, pseudoscalars, and tensor particles) can also be \textit{singly} produced (Fig.~\ref{fig:diags}, center). In all cases, $\gaga$ collisions present a very clean environment for measurements of processes with very few particles produced exclusively in the final state, very small or negligible irreducible backgrounds, and with the possibility, in the p-p case, to further constrain the collision kinematics with the simultaneous reconstruction of the momenta of the forward/backward $\gamma$-emitting protons in dedicated Roman Pots (RPs) detectors located inside the beamline~\cite{Piotrzkowski:2000rx,FP420RD:2008jqg,Tasevsky:2015xya,CMS:2021ncv}.

At the LHC, photon-photon interactions happen at unprecedentedly large effective luminosities at low masses in heavy-ion UPCs~\cite{Baltz:2007kq}, and up to very large $\gaga$ center-of-mass energies (up to a few~TeV) with UPCs with proton beams~\cite{deFavereaudeJeneret:2009db}. These facts have first revived the field of quantum electrodynamics (QED) at very high intensity initiated with the E-144 experiment at SLAC~\cite{E144:1996enr,Burke:1997ew}. In this context, the LHC has provided the first observation of light-by-light (LbL) scattering~\cite{dEnterria:2013zqi} (Fig.~\ref{fig:diags}, left) in lead-lead UPCs at the LHC, $\mathrm{Pb Pb} \overset{\gamma\gamma}{\to} \mathrm{Pb}\,\gamma\gamma\,\mathrm{Pb}$, at a nucleon-nucleon center of mass energy of $\sqrtsnn = 5.02$~TeV~\cite{ATLAS:2017fur,CMS:2018erd,ATLAS:2019azn}. Similarly, searches for LbL at the TeV scale have been carried out in pp at $\sqrts = 13$~TeV via $\mathrm{pp}\overset{\gamma \gamma}{\to} \mathrm{p}\,\gamma\gamma\,\mathrm{p}$ by tagging one or both protons in very forward RPs~\cite{TOTEM:2021zxa,CMS:2022zfd,ATLAS:2023zfc}. Such measurements have been used \eg\ to set competitive limits on nonlinear (Born--Infeld) extensions of QED~\cite{Ellis:2017edi}.

Photon-photon collisions at the LHC provide also very clean conditions for searches for particles beyond the standard model (BSM) that couple to photons~\cite{Baur:2001jj,Lietti:2002rq,Fichet:2014uka,Harland-Lang:2018iur,Schoeffel:2020svx,Shao:2022cly,dEnterria:2022sut}.
In particular, massive spin-0 particles, such as axion-like-particles (ALPs)~\cite{Knapen:2016moh,Baldenegro:2018hng,dEnterria:2021ljz,Goncalves:2021pdc}, as well as spin-2 tensor particles, such as gravitons~\cite{Atwood:1999zg,Ahern:2000jn,Inan:2012zz, Fichet:2014uka,Inan:2018jza,Shao:2022cly}, can be produced in photon-fusion processes (Fig.~\ref{fig:diags}, center), and manifest themselves as diphoton resonances on top of the LbL invariant mass continuum. Recent searches for excesses of exclusive diphotons produced above the LbL continuum~\cite{Knapen:2016moh} have allowed placing the most competitive limits on ALPs over masses $m_a\approx 5$--100~GeV in PbPb UPCs~\cite{CMS:2018erd,ATLAS:2020hii}, and over $m_a\approx 0.5$--2~TeV in pp collisions~\cite{TOTEM:2021zxa,CMS:2022zfd,ATLAS:2023zfc}.
Limits on ALP-photon coupling have also been set from searches for triphoton final states at electron-positron ($\epem$) colliders (Fig.~\ref{fig:diags}, right), using recent results from Belle~II and BES-III as well as from previous studies at LEP~\cite{Belle-II:2020jti,BESIII:2022rzz,Baillargeon:1995dg}. 

Whereas massive spin-0 particles have been extensively studied, the physics case for the two-photon production of massive spin-2 states at accelerators is still at an early 
stage, notwithstanding some exploratory works at colliders~\cite{Atwood:1999zg,Ahern:2000jn,Fichet:2014uka,Inan:2018jza,Shao:2022cly} and fixed-target facilities via the Primakoff process~\cite{Voronchikhin:2022rwc, Jodlowski:2023yne}.
In this paper, we extract new bounds on the photon-graviton coupling as a function of the graviton mass, by properly recasting the existing experimental searches for ALPs coupling to photons mentioned above. We do so by applying the experimental selection criteria to simulated ALP and graviton pseudodata generated within an effective field theory (EFT) approach, taking into account the different cross sections, diphoton partial widths, and decay kinematic distributions of the pseudoscalar and tensor particles, and using standard statistical methods.

Let us start by recalling that General Relativity (GR), as a classical field theory, describes the gravitational force in terms of an interacting massless tensor (spin-2) field. When the field is quantized, massless spin-2 particles, called gravitons, appear. The masslessness of the graviton is generally considered to be guaranteed by diffeomorphism invariance of GR~\cite{Hinterbichler:2011tt}. However, it is also known that gauge invariance does not always imply zero masses for gauge states. Quantum effects from other fields can, for example, give gravitons masses without breaking fundamental properties of GR. The fact that propagating degrees of freedom of gravity have mass is a fundamental issue with implications in many areas of physics including the propagation of gravitational waves~\cite{deRham:2016nuf}.
Also, the possible existence of a seflconsistent quantum field theoretical framework of GR valid at all energy scales is an open question~\cite{DeWitt:1967uc,DeWitt:1967yk}. 
Although such a theory remains elusive, one can however study practical and reliable consequences of the underlying quantum theory of GR by employing an EFT approach~\cite{Bjerrum-Bohr:2016hpa}.
New massive spin-2 degrees of freedom have been shown to arise in different modifications of gravity. Extradimensional theories of gravity, like the Arkani-Hamed–Dimopoulos–Dvali (ADD)~\cite{Arkani-Hamed:1998sfv} and Randall-Sundrum (RS)~\cite{Randall:1999ee} BSM models, proposed to explain the very large gap between the electroweak ($10^2$ GeV) and Planck ($10^{19}$ GeV) scales (“hierarchy problem”), generically predict massive tensor particles appearing as Kaluza–Klein (KK) excitations of these extra dimensions, either with a continuum mass spectrum, or as a number of widely separated resonances. Models with an extra dimension at the micron scale, predict also KK modes called ``dark gravitons'', that are a natural dark matter candidate \cite{Gonzalo:2022jac}. In addition, graviton-like particles, sometimes dubbed ``hidden gravitons''~\cite{Cembranos:2021vdv}, naturally appear in the context of bimetric theories of gravity \cite{Hassan:2011vm}.

In the following, we work under an EFT framework where the spacetime metric can be linearized and written in the form\footnote{It is worth noting that the separation of the metric into a background flat metric and a quantum perturbation $\kappa G_{\mu\nu}$, allows avoiding the conceptual problems of the standard interpretation of quantum mechanics applied to quantum gravity, i.e.\ that the classical observers doing preparations and measurements live themselves in the spacetime which they prepare and measure. Here, the observers and the experiment live in the classical, flat spacetime.}:
$g_{\mu\nu}=\eta_{\mu\nu} + \kappa G_{\mu\nu}$, with $G_{\mu\nu}$ the spin-2 quantum field (graviton) that we will assume to be potentially massive, and where $\kappa \sim 1/M_\mathrm{Pl}$ with $M_\mathrm{Pl}$ being the (dark/hidden) Planck mass. In this framework, the Einstein--Hilbert Lagrangian density takes the Fierz--Pauli expression, first formulated by them in 1939 to describe the linear theory of a massive spin-2 field~\cite{Fierz:1939ix}, and the interaction of the graviton with SM fields reads,
\be
\mathscr{L}^\mathrm{G}_{V,f} = \frac{k_{V,f}}{\Lambda} \ T_{\mu\nu}^{V,f} \ G^{\mu\nu}.
\label{eq:1}
\ee
Here, $k_{V,f}$ a factor that describes the strength of the coupling of the graviton field $G$ to the boson $V$ (including gauge and Higgs bosons) or fermion $f$, $\Lambda$ is an energy scale, and $T_{\mu\nu}^{V,f}$ is the energy-momentum tensor for bosons or fermions. 
In particular, for gravitons coupled to photons, the expression above gives:
\be
\mathscr{L}^\mathrm{G}_{\gamma} = g_{\mathrm{G}\gamma} \left(-F_{\mu\rho}F_\nu^{\rho}+\frac{1}{4}\eta_{\mu\nu} (F_{\rho\sigma})^2
\right) \ G^{\mu\nu},\;\mbox{with } g_{\mathrm{G}\gamma}\equiv \frac{k_{\gamma}}{\Lambda},
\label{eq:Lgraviton}
\ee
where $F_{\mu\rho}$ is the EM field, $\eta_{\mu\nu}$ the flat spacetime metric, and $g_{\mathrm{G}\gamma}$ is the G-$\gamma$ coupling.

In this work, we derive upper limits on $g_{\mathrm{G}\gamma}$ as a function of the graviton mass $m_\mathrm{G}$ using the experimental LHC and $\epem$ data mentioned above~\cite{CMS:2018erd,ATLAS:2023zfc,TOTEM:2021zxa,CMS:2022zfd,ATLAS:2023zfc,Belle-II:2020jti,BESIII:2022rzz,Baillargeon:1995dg} that correspond to probing $m_\mathrm{G}$ values from 100~MeV up to 2~TeV. We obtain these limits under two different scenarios. First, we take a simplified approach with a 100\% decay branching fraction of the graviton into two photons, $\mathcal{B}_{\mathrm{G} \to \gamma\gamma}=1$. 
Such a ``photophilic'' scenario is often assumed in ALPs searches~\cite{dEnterria:2021ljz,Agrawal:2021dbo}, and leads to a maximum sensitivity to the graviton-photon coupling. A second more realistic scenario is also considered with universal couplings of the graviton to all Standard Model (SM) particles~\cite{Das:2016pbk}. In this case, the graviton decay into diphotons is dominant only at low $m_\mathrm{G}$ values whereas above a few GeV, once the kinematic phase space for decays to massive SM fermions or bosons opens up, it amounts to $\mathcal{B}_{\mathrm{G} \to \gamma\gamma}\approx 0.05$. The universal couplings scenario also allows a proper computation of the $\epem \to \mathrm{G}\gamma$ cross sections 
without problems linked to violation of perturbative unitarity as described in Section~\ref{sec:ee_colls}.

The paper is organized as follows. The theoretical setup used to compute the graviton and ALP cross sections in photon-photon collisions at the LHC and in $\epem$ collisions is presented in Section~\ref{sec:th}. The generation and analysis of graviton and ALPs simulated samples, and the method of extraction of G-$\gamma$ coupling bounds from the experimental ALP limits, are discussed in Section~\ref{sec:analysis}. The derived limits as a function of $m_G$, including current bounds and future projections is presented in Section~\ref{sec:results}, together with their comparison to other existing results. The paper is closed with a summary in Section~\ref{sec:conclusion}.
 
%%%%%%%%%%%%%%%%%%%%% 
%\section{Position of the problem}
\section{Theoretical setup}
\label{sec:th}
%%%%%%%%%%%%%%%%%%%%%

The theoretical framework employed to study the production of gravitons and ALPs is presented, first, for photon fusion processes in UPCs at the LHC, $\gaga\to \mathrm{G},a\to\gaga$ (Fig.~\ref{fig:diags} middle), and via $\epem\to (\mathrm{G},a)\,\gamma \to 3\gamma$ final states (Fig.~\ref{fig:diags} right), second.

\subsection{Photon-photon collisions}
The description of the $\gaga\to \mathrm{G},a \to \gaga$ process is based on the EPA applied to ultrarelativistic protons or ions with low-virtuality equivalent photon fluxes, as implemented in the the \gammaUPC\ code~\cite{Shao:2022cly}. The cross section for the production of a given final state $X$ via photon fusion in an UPC of hadrons A and B with charges $Z_{1,2}$, $\mathrm{A B}\overset{\gamma\gamma}{\to} \mathrm{A} X \mathrm{B}$ can be written as a convolution integral of the product of the elementary cross section at a given $\gamma\gamma$ \cm\ energy, $\sigma_{\gamma\gamma\to X}(W_{\gamma\gamma})$, and the two-photon differential distribution of the colliding beams, 
\begin{eqnarray}
\sigma(\mathrm{A B}\overset{\gamma\gamma}{\to}\mathrm{A}X\mathrm{B})&=&\int{\frac{\mathrm{d}E_{\gamma_1}}{E_{\gamma_1}}\frac{\mathrm{d}E_{\gamma_2}}{E_{\gamma_2}}\frac{\mathrm{d}^2N^{(\mathrm{AB})}_{\gamma_1/Z_1,\gamma_2/Z_2}}{\mathrm{d}E_{\gamma_1}\mathrm{d}E_{\gamma_2}}\sigma_{\gamma\gamma\to X}(W_{\gamma\gamma})},
\label{eq:sigma_gaga1}
\end{eqnarray}
where $W_{\gamma\gamma}^2=4E_{\gamma_1}E_{\gamma_2}$ is the \cm\ energy of the collision of photons with energies $E_{\gamma_1}$ and $E_{\gamma_2}$, and 
\begin{eqnarray}
\frac{\mathrm{d}^2N^{(\mathrm{AB})}_{\gamma_1/Z_1,\gamma_2/Z_2}}{\mathrm{d}E_{\gamma_1}\mathrm{d}E_{\gamma_2}}&=&\int{\mathrm{d}^2\mathbf{b}_1\mathrm{d}^2\mathbf{b}_2P_\mathrm{no~inel}(\mathbf{b}_1,\mathbf{b}_2)N_{\gamma_1/Z_1}(E_{\gamma_1},\mathbf{b}_1)N_{\gamma_1/Z_1}(E_{\gamma_2},\mathbf{b}_2)},
\end{eqnarray}
is the effective two-photon luminosity accounting for the probability $P_\mathrm{no~inel}(\mathbf{b}_1,\mathbf{b}_2)$ of hadrons A and B to remain intact after their interaction. In the expressions above, $N_{\gamma_i/Z_i}(E_{\gamma_i},\mathbf{b}_i)$ is the photon number density with the photon energy $E_{\gamma_i}$ at the impact parameter $\mathbf{b}_i$ from the $i$th initial hadron.
The photon number densities are usually derived from two different hadron form factors, such as the electric-dipole (EDFF, Eq.~(11) in~\cite{Shao:2022cly}) and charge (ChFF, Eq.~(13) in~\cite{Shao:2022cly}) form factors. In the EDFF case, because the photon number density is divergent at low values of the impact parameter $b\equiv \left|\mathbf{b}\right|$, arbitrary $b_1>R_A$ and $b_2>R_B$ cuts must be imposed, with $R_{A,B}$ being the radii of hadrons A and B. On the other hand, such an issue is absent in the ChFF case, and one can safely integrate $b_{1,2}$ down to zero. Although the $\gaga$ cross sections obtained with EDFF and ChFF fluxes are in general similar, the ChFF is a more realistic, and therefore preferable, choice. In the latter formula, we have integrated over the virtualities $Q^2$ of the initial photons, which can be certainly unintegrated in order to make explicit their very small values, typically of order 
 $Q^2\sim R_A^{-2}\lesssim 0.08$~GeV$^2$ for protons ($R_\mathrm{p}\approx 0.8$~fm), and $Q^2\lesssim 10^{-3}$~GeV$^2$ for Pb nuclei ($R_\mathrm{A}\approx 7$~fm)~\cite{Shao:2022cly}.

In the case of heavy-ion beams, the action of all the charges in the nucleus adds coherently and the photon flux is enhanced by a $Z^2$ factor compared to the proton case, leading to a $Z_1^2Z_2^2$ increase in the corresponding $\gaga$ cross sections. The nonoverlap hadronic interaction probability density $P_{\mathrm{no~inel}}(\mathbf{b}_1,\mathbf{b}_2)$ depends on the spatial separation of the two initial hadrons, i.e., $P_{\mathrm{no~inel}}(\mathbf{b}_1,\mathbf{b}_2)=P_{\mathrm{no~inel}}(\left|\mathbf{b}_1-\mathbf{b}_2\right|)$, and can be derived from the standard opacity (optical density) computed from realistic hadronic transverse profile overlap functions with a Glauber Monte Carlo (MC) model~\cite{dEnterria:2020dwq}.

The expected LbL continuum cross sections can be calculated through Eq.~(\ref{eq:sigma_gaga1}) plugging in the elementary $\gaga\to\gaga$ cross section and using a proper setup for the photon fluxes and nonoverlap probabilities. For the resonant graviton and ALP total cross sections, a more convenient equation can be employed. The cross section for the exclusive production of a C-even resonance $X$ of spin $J$ and two-photon decay width $\Gamma_{\gaga}(X)$, through $\gaga$ fusion in an UPC of charged particles A and B, 
reads now~\cite{Budnev:1975poe}
\begin{equation}
\sigma(\mathrm{A}\; \mathrm{B}\,\xrightarrow{\gaga} \mathrm{A} \; X \; \mathrm{B}) = 
4\pi^2 (2 J+1)\frac{\Gamma_{\gaga}(X)}{m_X^2} 
  \left. \frac{\mathrm{d}{\Lumi}^{(\mathrm{A}\,\mathrm{B})}_{\gaga}}{\mathrm{d}W_{\gaga}} \right|_{W_{\gaga}=m_X},
\label{eq:sigma_AA_X}
\end{equation}
where $\frac{d{\Lumi}^{(\mathrm{A\,B})}_{\gaga}}{dW_{\gaga}}\big|_{W_{\gaga}=m_X}$ is the value of the effective two-photon luminosity at the resonance mass $m_X$ in an UPC at nucleon-nucleon \cm\ energy $\sqrtsnn$, and amounts to
\begin{eqnarray}
\frac{\mathrm{d}{\Lumi}^{(\mathrm{AB})}_{\gaga}}{\mathrm{d}W_{\gaga}}&=&\frac{2W_{\gaga}}{s_{_\mathrm{NN}}}\int{\frac{\mathrm{d}E_{\gamma_1}}{E_{\gamma_1}}\frac{\mathrm{d}E_{\gamma_2}}{E_{\gamma_2}}\delta\left(\frac{W_{\gaga}^2}{s_{_\mathrm{NN}}}-\frac{4E_{\gamma_1}E_{\gamma_2}}{s_{_\mathrm{NN}}}\right)\frac{\mathrm{d}^2N^{(\mathrm{AB})}_{\gamma_1/\mathrm{Z}_1,\gamma_2/\mathrm{Z}_2}}{\mathrm{d}E_{\gamma_1}\mathrm{d}E_{\gamma_2}}}\,.\label{eq:gagalumi}
\end{eqnarray}

From Eq.~(\ref{eq:sigma_AA_X}), one can straightforwardly see that, for the same values of resonance masses and diphoton partial decay widths, the photon-fusion production of gravitons ($J=2$) will be enhanced by a factor of $(2 J+1) = 5$ compared to the ALPs ($J=0$) case. Such an apparent benefit will be, however, outplayed by a comparatively reduced graviton coupling to photons, as explained below.
The calculation of their expected photon-fusion cross sections through Eq.~(\ref{eq:sigma_AA_X}) relies on computing their $\Gamma_{\gaga}(X)$ two-photon widths with a given interaction Lagrangian.
For the ALP case, 
 $\gamma \gamma \to a \to \gamma \gamma$, the relevant Lagrangian is
\begin{eqnarray}
\mathscr{L}&\supset& \frac{1}{2}\partial_\mu a \partial^\mu a-\frac{m_a^2}{2}a^2-\frac{g_{a\gamma}}{4}a F^{\mu \nu}\tilde{F}_{\mu\nu},\;\mbox{with } g_{a\gamma} \equiv C_{\gaga}/\Lambda, 
\label{eq:Laxion}
\end{eqnarray}
where $a$ is the ALP field, $\tilde{F}_{\mu\nu}$ is the photon field strength dual tensor, and the dimensionful ALP-$\gamma$ coupling strength $g_{a\gamma}$ is inversely proportional to the high-energy scale $\Lambda$ associated with the spontaneous breaking of an approximate Peccei–-Quinn global U$(1)$ symmetry~\cite{Peccei:1977hh}, and the effective dimensionless coefficient $C_{\gaga}$ rescales the ALP-$\gamma$ coupling whenever the ALP also interacts with (and, therefore, decays to) other SM particles (although most often the photon-dominance, or photophilic $C_{\gaga} = 1$ case is considered in the literature)~\cite{Agrawal:2021dbo}.

The production cross sections for massive gravitons via $\gamma \gamma \to \mathrm{G} \to \gamma \gamma$ can be similarly obtained from the Fierz--Pauli Lagrangian, Eq.~(\ref{eq:Lgraviton}). Writing explicitly the kinetic content for the graviton field of mass $m_\mathrm{G}$, it reads:
\be
\mathscr{L}_\mathrm{FP} = -\frac{1}{2} (\partial_\rho G_{\mu\nu})^2 + \partial_\mu G_{\nu\rho} \partial^\nu G^{\mu\rho}
-\partial_\mu G^{\mu \nu} \partial_\nu G + \frac{1}{2} (\partial_\rho G)^2 
-\frac{1}{2}m_\mathrm{G}^2 \left( (G_{\mu\nu})^2-G^2\right),
\ee
from which the propagator for the graviton field, represented by the dotted line in Fig.~\ref{fig:diags} (center), can be computed directly as
\be
T^{\mu\nu\rho\sigma} = \frac{i}{p^2-m_\mathrm{G}^2+i\epsilon} \left(
\frac{1}{2} (P_{\mu\rho}P_{\nu\sigma}+P_{\mu\sigma}P_{\nu\rho}) - \frac{1}{3} P_{\mu\nu} P_{\rho\sigma} \right),
\label{eq:LFP}
\ee
with $P_{\mu\nu} = \eta_{\mu\nu} + p_\mu p_\nu/m_\mathrm{G}^2$. 
In this latter expression we see the $m_\mathrm{G}$ pole in mass that gives the resonant effect in the invariant mass LbL spectrum\footnote{Let us note that in the massless case, the structure of the propagator is preserved, but with some modifications. For a massless graviton, Eq.~(\ref{eq:LFP}) would give:
\be
T^{\mu\nu\rho\sigma}_{m=0} = \frac{1}{2} \frac{i}{p^2+i\epsilon} \left(
 \eta_{\mu\rho}\eta_{\nu\sigma}+\eta_{\mu\sigma}\eta_{\nu\rho} - \eta_{\mu\nu} \eta_{\rho\sigma} \right),
 \label{eq:prop2}
\ee
which,\,interestingly,\,does not lead to a resonant effect but a particular behavior of the cross section in the forward~limit.}.

The generation of ALP and graviton simulated events in this work is carried out with the \gammaUPC\ code~\cite{Shao:2022cly}, using ChFF $\gamma$ fluxes for protons and ions and computing the nonoverlap probabilities with a Glauber MC~\cite{Loizides:2017ack}, combined with \madgraph~\cite{Alwall:2014hca,Frederix:2018nkq} (hereafter identified as \mgshort) where the corresponding Lagrangians, Eqs.~(\ref{eq:Lgraviton}) and (\ref{eq:Laxion}), are coded as input models in the Universal Feynman Output (\ufo) format~\cite{Degrande:2011ua,Darme:2023jdn}.
We have compared the computed cross section for ALP or graviton production with the results of several alternative codes~\cite{Fichet:2014uka, Alwall:2014hca,Shao:2022cly} finding fully consistent results (and, thus, also the corresponding graviton exclusion limits).

\subsection{Electron-positron collisions}
\label{sec:ee_colls}

We consider next the graviton and ALP production cross sections in $\epem$ collisions through the process shown in Fig.~\ref{fig:diags} (right), and describing their photon couplings with the same Lagrangians, Eqs.~(\ref{eq:Lgraviton}) and (\ref{eq:Laxion}) respectively, used for photon-photon collisions. For $\epem$ collisions at Belle~II and LEP energies, the leading-order inclusive cross section, neglecting the tiny electron mass $m_e$, reads
\begin{eqnarray}
\sigma(\epem\to a \gamma \to \gamma\gamma\gamma)=\frac{\alpha g_{a\gamma}^2}{24}\frac{(s-m_a^2)^3}{s^3}\;\mathcal{B}_{a\to \gamma\gamma},\;\text{for ALPs, and}
\label{eq:sigma_ee_ALP}
\end{eqnarray}
\begin{eqnarray}
\sigma(\epem\to \mathrm{G} \gamma \to \gamma\gamma\gamma) = \frac{\alpha}{36}
\left(\frac{k_{\gamma}}{\Lambda}\right)^2\frac{(s-m_\mathrm{G}^2)^3}{s^3}\frac{s^2+3s m_\mathrm{G}^2+6m_\mathrm{G}^4}{m_\mathrm{G}^4}\;\mathcal{B}_{\mathrm{G}\to \gamma\gamma},\;\text{for gravitons,}
\end{eqnarray}
where $s$ is the squared center-of-mass energy of the collision, and $\mathcal{B}_{a,\mathrm{G}\to \gamma\gamma}$ the corresponding $a,\mathrm{G}\to \gamma\gamma$ branching fractions. This latter expression indicates that the graviton cross section, as opposed to the ALP one, has the asymptotic form
\begin{eqnarray}
  \lim_{s\gg m_\mathrm{G}^2}{\sigma(\epem\to \mathrm{G} \gamma \to \gamma\gamma\gamma)}&=&\frac{\alpha }{36}
  \left(\frac{ k_{\gamma}}{\Lambda}\right)^2\frac{s^2}{m_\mathrm{G}^4}\;\mathcal{B}_{\mathrm{G}\to \gamma\gamma},
\end{eqnarray}
which is divergent in the $m_\mathrm{G}^2/s\to 0$ limit. Such a unitarity-violating behavior is due to the assumption that the graviton couples only to photons, and not to electrons. A more realistic universal-coupling scenario for gravitons can solve this perturbative unitarity problem~\cite{Das:2016pbk,Gill:2023kyz}. In such a universal-coupling scenario, the expression for the $\epem\to \mathrm{G} \to 3\gamma$ cross section, reads
\begin{eqnarray}
\sigma(\epem\to \mathrm{G} \gamma \to \gamma\gamma\gamma)&=&\frac{\alpha }{24}\left(\frac{ k_U}{ \Lambda}\right)^2\frac{(s-m_\mathrm{G}^2)^3}{s^3}\;\mathcal{B}_{\mathrm{G}\to \gamma\gamma},
\label{eq:belle}
\end{eqnarray}
which, as its ALPs counterpart given by Eq.~(\ref{eq:sigma_ee_ALP}), is now well-behaved for all $m_\mathrm{G}$.

\begin{figure}[!htbp]
\centering
  \includegraphics[width=0.85\textwidth]{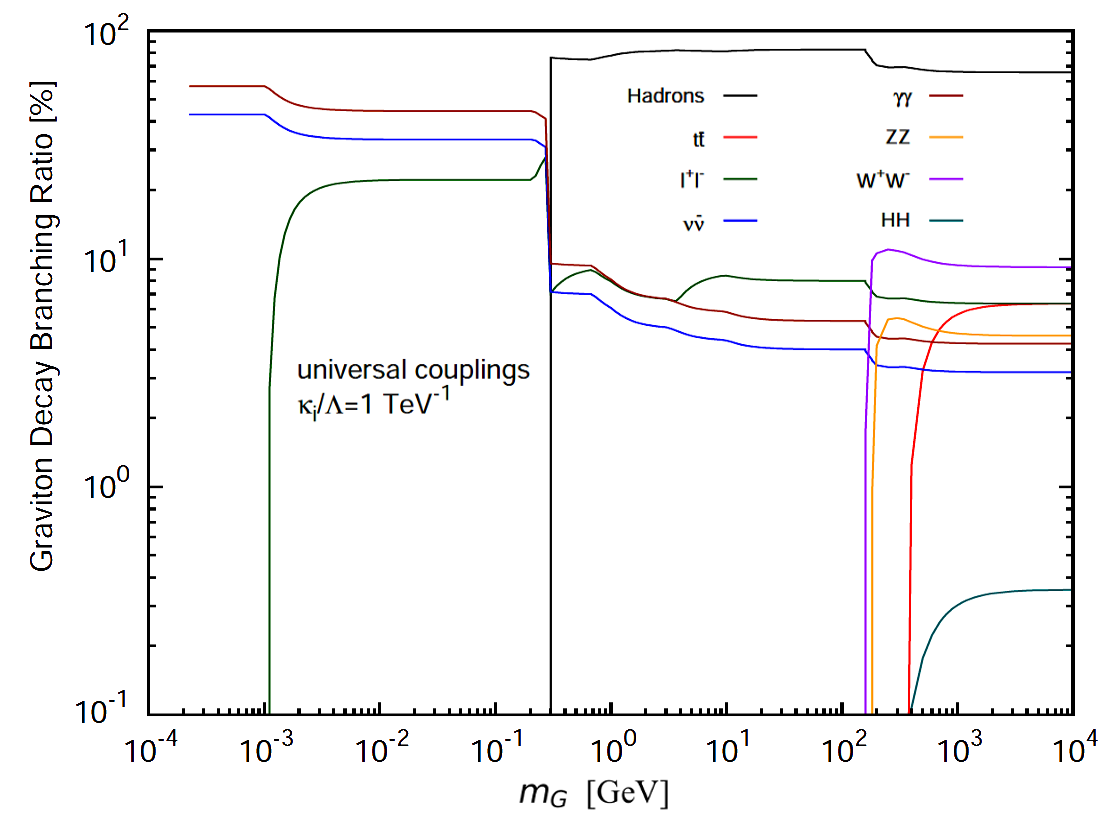}
 \caption{Branching ratios for the various decay modes of a massive graviton as a function of its mass $m_{\mathrm{G}}$ assuming its universal-coupling with the SM particles. Specific $\mathcal{B}_{\mathrm{G}\to XX}$ numerical values are given in Table~\ref{tab:branching_ratios}.}
 \label{fig:diagbr}
\end{figure}

Of course, allowing for other couplings reduces also the diphoton decay probability for massive gravitons. In principle, for the graviton production in $\epem$ collisions via the diagram shown in Fig.~(\ref{fig:diags}) (right), one could just consider a simplified model with universal couplings to photons and electrons alone, $k_{U}=k_{\gamma}=k_{e}$, neglecting all other couplings. In this case, the asymptotic cross section for $s\gg m_\mathrm{G}^2$ can be written as: $\sigma \approx \frac{\alpha }{6}(\frac{ k_{U}}{\Lambda})^2\mathcal{B}_{\mathrm{G}\to \gamma\gamma}$, and the two partial widths would be: $\Gamma(\mathrm{G}\to\gamma\gamma)=(\frac{k_\gamma}{\Lambda})^2\frac{m_\mathrm{G}^3}{80\pi}$ and $\Gamma(\mathrm{G}\to\epem)=(\frac{k_e}{\Lambda})^2\frac{m_\mathrm{G}^3}{160\pi}(1-\frac{4m_e^2}{m_\mathrm{G}^2})^{3/2}(1+\frac{8m_e^2}{3m_\mathrm{G}^2})$. Asymptotically, one would then have $\mathcal{B}_{\mathrm{G}\to \gamma\gamma}=\frac{2}{3}$ when $m_\mathrm{G}\gg 2m_e$, and only when $m_\mathrm{G}\lesssim 2m_e$ the diphoton branching fraction would be unity. This simple example shows that for the range of graviton masses probed by the Belle~II and LEP data ($m_\mathrm{G}\approx 0.1$--100~GeV), the assumption of $\mathcal{B}_{\mathrm{G}\to \gamma\gamma}=1$ would be incorrect.
The actual decay branching fractions of the graviton to all SM particle pairs as a function of $m_\mathrm{G}$ in the universal-couplings scenario are shown in Fig.~\ref{fig:diagbr}, and Table~\ref{tab:branching_ratios} collects a few reference values as a guideline. One can see now that the diphoton decay is relatively dominant only in the case of gravitons with masses below twice the pion mass ($m_\mathrm{G}\lesssim 0.25$~GeV) with values $\mathcal{B}_{\mathrm{G}\to \gamma\gamma}\approx 40\%$, whereas hadronic decays take over for heavier gravitons. Above $m_\mathrm{G}\approx 5$~GeV, the diphoton decay amounts to $\mathcal{B}_{\mathrm{G}\to \gamma\gamma}\approx 5\%$, which would at face value translate into factors of $\sim$20 less constraining limits placed on gravitons compared to ALPs searches in the photon-dominance assumption often consider for the latter ($C_{\gaga}=1$ in Eq.~(\ref{eq:Laxion}) leading to $\mathcal{B}_{a\to \gamma\gamma} = 1$).

%%%%%%%%%%%%%%%%%%%%%%%
%%%%%%%%%%%%%%%%%%%%%%%
%%%%%%%%%%%%%%%%%%%%%%%
\begin{table}[htbp!]
\centering
\caption{Branching ratios for various decay modes of a graviton of different masses, assuming its universal coupling with SM particles.
\label{tab:branching_ratios}}
\begin{tabular}{cccccc}
\hline
\multicolumn{6}{c}{Graviton decay ${\mathcal B}_{\mathrm{G}\to XX}$(\%)} \\ \hline
{Channel} & {$m_{\mathrm{G}}=100$~MeV} & {$m_{\mathrm{G}}=5$~GeV} & {$m_{\mathrm{G}}=100$~GeV} & {$m_{\mathrm{G}}=1$~TeV} & {$m_{\mathrm{G}}=2$~TeV} \\ \hline
$\gamma$ $\gamma$ & 44.4 & 6.1 & 5 & 4.3 & 4.2\\ %\hline
$\nu$ $\bar{\nu}$ & 33.3 & 4.5 & 4 & 3.2 & 3.2 \\ %\hline
$l^{+}l^{-}$ & 22.2 & 7.8 & 8 & 6.4 & 6.4 \\ %\hline
Hadrons & -- & 81.5 & 82 & 66 & 65.8\\ %\hline
ZZ & --& -- & -- & 4.7 & 4.6\\ %\hline
W$^{+}$W$^{-}$ & -- & -- & -- & 9.4 & 9.2 \\ %\hline
HH & -- & -- & -- & 0.3 & 0.4\\ %\hline
$\mathrm{t}\bar{\mathrm{t}}$ & -- &-- & -- & 5.7 & 6.2\\ \hline
\end{tabular}
\end{table}
%%%%%%%%%%%%%%%%%%%%%%%

At BES-III, the underlying production process differs from the Belle~II and LEP cases as the $a,$\,G resonance is not directly radiated from the $s$-channel (virtual) $\gamma^*$ or Z$^*$ boson (Fig.~\ref{fig:diags}, right), but an intermediate $\jpsi$ meson is first produced that decays into the ALP or graviton plus a photon, leading to the three-photon final state. At leading order, the partial width of the $J/\psi\to a\gamma\to \gamma\gamma\gamma$ decay reads
\begin{eqnarray}
\Gamma(J/\psi\to a\gamma\to\gamma\gamma\gamma)&=&\frac{\alpha }{81}g_{a\gamma}^2\left(1-\frac{m_{a}^2}{m_{J/\psi}^2}\right)^3\langle O^{J/\psi}\rangle\;\mathcal{B}_{a\to \gamma\gamma},
\label{eq:GammaJpsi_agamma}
\end{eqnarray}
where $\langle O^{J/\psi}\rangle$ is the long-distance matrix element of the $J/\psi$ particle. For the graviton production, 
a photon-only coupling will lead to the same perturbative unitarity violation problem mentioned above, and we have to work in the universal coupling scenario. In this case, the leading order partial width of $J/\psi\to \mathrm{G} \gamma\to \gamma\gamma\gamma$ is given by:
\begin{eqnarray}
\!\!\!\!\!\!\Gamma(J/\psi\to \mathrm{G} \gamma\to\gamma\gamma\gamma)&=&\frac{2\alpha}{243} \left(\frac{k_U}{\Lambda}\right)^2 \left(1-\frac{m_\mathrm{G}^2}{m_{J/\psi}^2}\right)\left(1+3\frac{m_\mathrm{G}^2}{m_{J/\psi}^2}+6\frac{m_\mathrm{G}^4}{m_{J/\psi}^4}\right) \langle O^{J/\psi}\rangle\;\mathcal{B}_{\mathrm{G}\to \gamma\gamma}.
\label{eq:GammaJpsi_Ggamma}
\end{eqnarray}
Combining Eqs.~(\ref{eq:GammaJpsi_agamma}) and (\ref{eq:GammaJpsi_Ggamma}), we can derive a bound on the G-$\gamma$ coupling from any given one obtained for the ALP-$\gamma$ case via
\begin{eqnarray}
\left(\frac{k_U}{\Lambda}\right)&=&\big(g_{a\gamma}\big)\;\frac{\left(m_{J/\psi}^2-m_\mathrm{G}^2\right)}{\sqrt{\left(4m_\mathrm{G}^4+2m_\mathrm{G}^2m_{J/\psi}^2+\frac{2}{3}m_{J/\psi}^4\right)\;\mathcal{B}_{\mathrm{G}\to \gamma\gamma}}},
\label{eq:limitGaFromJpsi}
\end{eqnarray}
where we have assumed $\mathcal{B}_{a\to \gamma\gamma}=1$.\\

For $\epem$ collisions, the generation of simulated graviton and ALPs events is performed with \mgshort, with the universal-couplings setup of Ref.~\cite{Das:2016pbk} for the graviton case and using the Lagrangian Eq.~(\ref{eq:Laxion}) for the ALP samples, coded both also in the \ufo\ format.

%%%%%%%%%%%%%%%%%%%%
%\section{Analysis strategy and detector effects}
\section{Analysis of the simulated data} %and experimental data}
\label{sec:analysis}
%%%%%%%%%%%%%%%%%%%%

Simulated events are generated using the theoretical setup discussed in the previous section, for all ALP and graviton production processes at the LHC and in $\epem$ collisions at BES-III and Belle~II\footnote{LEP bounds are less competitive than the LHC ones, and event samples are not explicitly generated for them.} listed in Table~\ref{tab:Graviton} for the relevant mass ranges.

%%%%%%%%%%%%%%%%%%%%%%%
\begin{table}[htbp!]
\centering
\caption{Summary of the six ALP and graviton production processes considered in this work, along with the mass ranges experimentally probed.}
\label{tab:Graviton}
\begin{tabular}{lccc}
\hline
Process & Colliding system & nucleon-nucleon or $\epem$ \cm\ energy & $m_{a,\mathrm{G}}$ range \\
\hline
$\gamma\gamma \to a,\mathrm{G} \to \gamma\gamma$ & PbPb & 5.02 TeV & 5--100 GeV\\
$\gamma\gamma \to a,\mathrm{G} \to \gamma\gamma$ & pp & 14 TeV & 0.15--2~TeV\\
$a,\mathrm{G}\; \gamma \to \gamma\gamma\; \gamma $ & $\epem$ & 3--11~GeV & 0.16--10 GeV\\
\hline
\end{tabular}
\end{table}

\begin{figure}[!htbp]
\centering
 \includegraphics[width=0.67\textwidth]{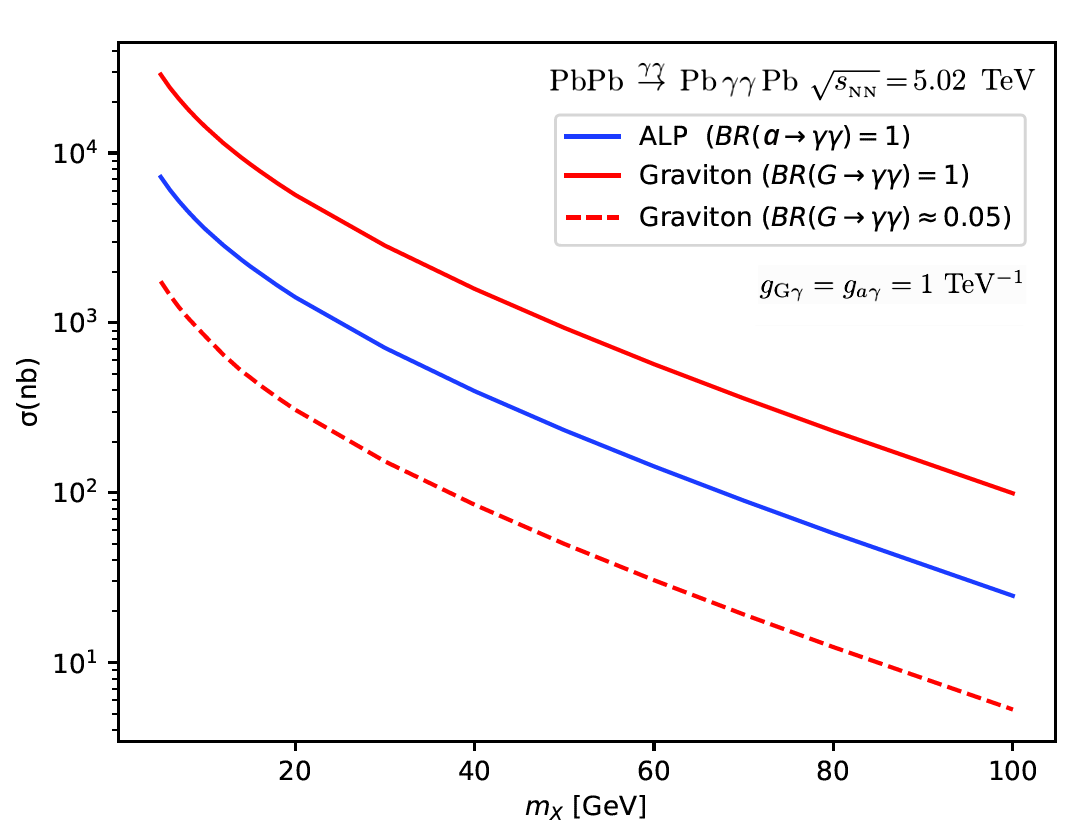}
 \caption{Total $\gaga$ cross sections for graviton and ALP production in PbPb UPCs at 5.02~TeV as a function of resonance mass, for the same photon couplings values, $g_{\mathrm{G} \gamma} = g_{a \gamma} = 1$~TeV$^{-1}$, and two different assumptions (photophilic or universal) on the graviton-photon coupling.}
 \label{fig:pbpbxs}
\end{figure}

As an example, Fig.~\ref{fig:pbpbxs} shows the computed total cross sections for graviton and ALP production versus mass in PbPb UPCs at $\sqrtsnn=5.02$~TeV, for the same photon-coupling values $g_{\mathrm{G} \gamma} = g_{a \gamma} = 1$~TeV$^{-1}$. The general trend for both particles is similar, featuring a decrease of the cross sections as a function of mass due to the $\sigma\propto m_{X}^{-2}$ dependence of Eq.~(\ref{eq:sigma_AA_X}) and the reduced effective $\gaga$ luminosity $\frac{\mathrm{d}{\Lumi}^{(\mathrm{A}\,\mathrm{B})}_{\gaga}}{\mathrm{d}W_{\gaga}} \big|_{W_{\gaga}=m_X}$ for increasing $W_{\gaga}$ \cm\ energy. Assuming $\mathcal{B}_{a,\mathrm{G}\to\gamma\gamma}=1$, one can observe graviton production cross sections (solid red curve) about five times larger than the ALP ones (blue solid curve), as given from the different spin counting of the two particles in Eq.~(\ref{eq:sigma_AA_X}). However, considering the more realistic scenario of universal couplings for the graviton, $\mathcal{B}_{\mathrm{G}\to\gamma\gamma}\approx 0.05$ (dashed red curve), and keeping the photon-dominance case for the ALP, we see that the final cross sections for $\mathrm{Pb Pb} \overset{\gamma\gamma}{\to} \mathrm{Pb}\,X(\gamma\gamma)\,\mathrm{Pb}$ are about four times smaller for gravitons than for ALPs.
%%%%%%%%%%%%%%%%%%%%%%%

For the extraction of upper limits on the photon-fusion graviton cross sections and, thus, on the $g_{\mathrm{G}\gamma}$ 
coupling, one can proceed along two different but equivalent approaches:
\begin{description}
 \item (i) One can use full simulations for the production cross sections for gravitons and associated backgrounds, applying the same requirements as in the experimental analyses, accounting for all detector effects, and employing a standard statistical framework for limits setting based on the experimental results and the generated pseudodata, as described in~\cite{Schoeffel:2020svx,Baldenegro:2018hng}.
 \item (ii) Or else, one can use the existing ALPs limits derived from the data, and properly reinterpret them for the graviton case, taking into account all differences between the production and decay properties of both BSM particles after applying all experimental analysis selection criteria.
\end{description}
We employ here the second technique, and show in the Appendix~\hyperref[sec:App]{A} its statistical equivalence to the first method. In the approach (ii), the $\sigma_{a}$ cross section for ALP production can be derived from the interacting Lagrangian, Eq.~(\ref{eq:Laxion}), and is proportional to $g_{a \gamma}^2 \times \mathcal{B}_{a \to \gamma \gamma}$. Similarly, following Eq.~(\ref{eq:Lgraviton}), the cross section for gravitons, $\sigma_\mathrm{G}$, is proportional to $g_{\mathrm{G} \gamma}^2 \times \mathcal{B}_{\mathrm{G} \to \gamma \gamma}$. Then, any bound obtained for the ALP-$\gamma$ coupling at a given $m_{\gaga}$ bin can be converted into the corresponding bound for the G-$\gamma$ coupling via
\be
g_{\mathrm{G} \gamma} = \sqrt{\frac{\sigma_{a}}{\sigma_{\mathrm{G}}}} \times \frac{\mathcal{A}_\mathrm{G}}{\mathcal{A}_\mathrm{a}} \times g_{a \gamma}.
\label{eq:recast}
\ee
Here $\mathcal{A}_{a}/\mathcal{A}_\mathrm{G}$ is the ratio of experimental fiducial acceptances for ALPs and gravitons decaying into a pairs of photons. Tensor particles decay on average into softer and more isotropic photons than pseudoscalar particles. This latter factor is derived from our full simulations, after applying the fiducial criteria of each experiment, and amounts to about a 10\% (50\%) correction at high (low) masses. The same formula can be used to set graviton limits from those placed on ALPs at the Belle~II and LEP experiments. The case of BES-III is slightly different, and the graviton limits are directly obtained through Eq.~(\ref{eq:limitGaFromJpsi}). In the Appendix~\hyperref[sec:App]{A}, a proof of the equivalence between both techniques (i) and (ii) is given. In particular, we demonstrate that a graviton search limit based on method (i) implies Eq.~(\ref{eq:recast}). 

In order to obtain the final limits on $g_{\mathrm{G} \gamma}$ through Eq.~(\ref{eq:recast}), we need to implement all experimental analyses and apply on our simulated samples the same selection requirements applied for ALP searches in the data. 
The searches carried out in PbPb UPCs are currently the most competitive for ALPs in the range $m_a \approx 5$--100~GeV. In this case, the final state of interest involves the observation of two exclusive photons with transverse energy $E_{T} \gtrsim 2$ GeV, emitted over $\lvert \eta \rvert\lesssim 2.4$ pseudorapidities, and pair invariant masses exceeding 5~GeV, with a rapidity gap requirement of no other significant hadronic activity occurring within $\lvert \eta \rvert< 5$. To further refine the analysis and reduce background contamination, additional kinematic criteria are applied to the photon pair, including selections on diphoton transverse momentum ($\pT^{\gamma\gamma}$) below 1~GeV, and on acoplanarity ($A_{\phi}^{\gamma\gamma}\equiv 1-|\Delta \phi_{\gamma \gamma}|/\pi$) less than $\approx 0.01$. These two additional criteria enhance the sensitivity to photon-fusion production processes that are characterized by the production of a central system at rest that decays into two photons in a back-to-back configuration, while minimizing contributions from misidentified $\gamma \gamma \to \epem (\gamma,\gaga)$ events. The full list of requirements applied to our simulated data to reproduce the ATLAS and CMS measurements are summarized in Table~\ref{tab:observable-value-combined}.

%%%%%%%%%%%%%%%%%%%%%%%
\tabcolsep=4.mm
\begin{table}[htbp!]
\centering
\caption{Selection criteria applied in the analyses of simulated ALP and graviton samples, following the ATLAS and CMS measurements of exclusive diphotons in PbPb~\cite{ATLAS:2020hii,CMS:2018erd} and pp~\cite{CMS:2022zfd,ATLAS:2023zfc}~UPCs.
}
\label{tab:observable-value-combined}
\begin{tabular}{l|cc|cc}\hline
Variable & \multicolumn{2}{c}{$\mathrm{Pb Pb} \overset{\gamma\gamma}{\to} \mathrm{Pb}\,\gamma\gamma\,\mathrm{Pb}$} & \multicolumn{2}{c}{$\mathrm{pp}\overset{\gamma \gamma}{\to} \mathrm{p}\,\gamma\gamma\,\mathrm{p}$}\\
 & (ATLAS) & (CMS) & (ATLAS) & (CMS) \\\hline
$\sqrtsnn$ \cm\ energy (TeV) & 5.02 & 5.02 & 13.0 & 13.0 \\ %\hline
Integrated luminosity $\mathcal{L}$ & 2.2 nb$^{-1}$ & 0.4 nb$^{-1}$ & 14.6 fb$^{-1}$ & 9.4 fb$^{-1}$\\
Exclusive number of photons & 2 & 2 & 2 & 2 \\ %\hline
%$E^{\gamma}_{T}$ 
Single photon $\pT^\gamma$ & $>2.5$ GeV & $>2$ GeV & $>40$ GeV & $>100$ GeV \\%\hline
Single photon $|\eta^\gamma|$ & $< 2.37$ & $< 2.4$ & $< 2.37$ & $< 2.5$ \\
Pair $\pT^{\gamma\gamma}$ & $<1$ GeV & $<1$ GeV & $<1$ GeV & $<1$ GeV \\
Pair $m_{\gamma \gamma}$ & $> 5$ GeV & $> 5$ GeV & $> 150$ GeV & $> 200$ GeV \\
Pair acoplanarity $A^{\gamma \gamma}_{\phi}$ & $<0.01$ & $<0.01$ & $<0.01$ & $<0.01$ \\%\hline
Rapidity gap range $|\eta^\mathrm{gap}|$ & $< 5$ & $< 5$ & -- & -- \\
Proton tagging & -- & -- & single & double \\
Proton energy loss $\xi$ & -- & -- & [0.035--0.08] & [0.02--0.2] \\
%\hline
\hline
\end{tabular}
\end{table}
%%%%%%%%%%%%%%%%%%%%%%%

\begin{figure}[!htbp]
\centering
 \includegraphics[width=0.65\textwidth]{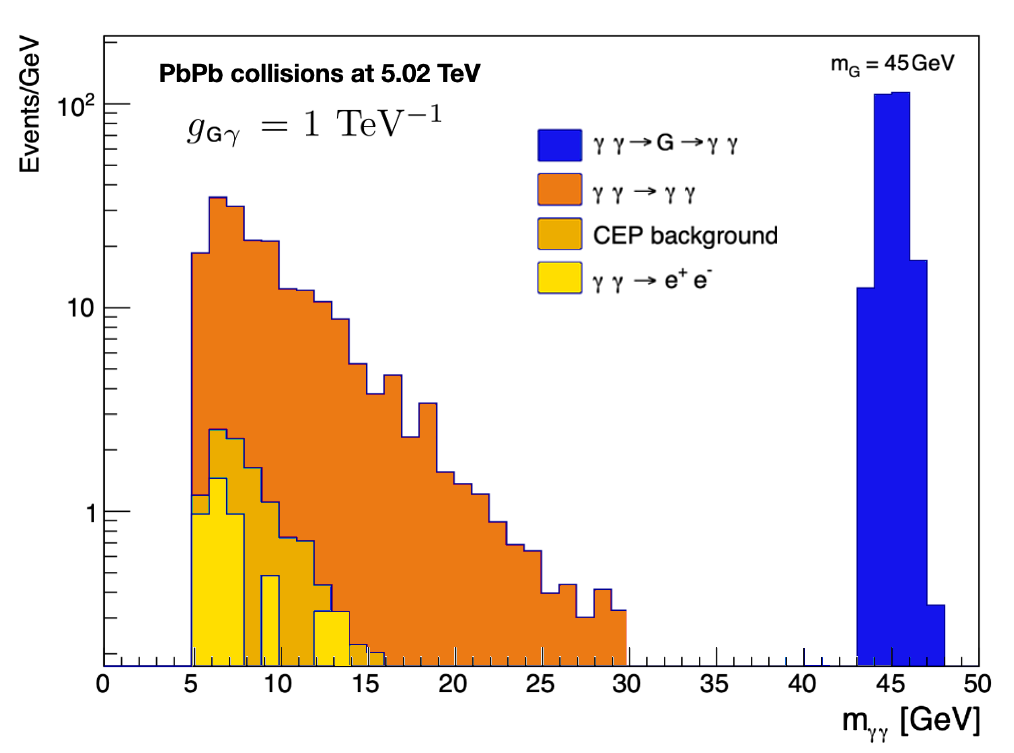}
 \caption{Simulated invariant mass distribution of exclusive photon pairs produced in PbPb UPCs at 5.02~TeV for a graviton signal ($m_\mathrm{G} = 45$~GeV mass and $g_{\mathrm{G}\gamma}=1$~TeV$^{-1}$ coupling), and LbL scattering (orange), CEP, and $\gaga\to\epem$ background processes (dark and light yellows). All distributions are presented with an emulation of diphoton detector resolution and inefficiencies.}
 \label{fig:massplot}
\end{figure}

Figure~\ref{fig:massplot} shows a typical diphoton invariant mass distribution for a generated graviton signal (with $m_\mathrm{G} = 45$~GeV and $g_{\mathrm{G}\gamma}=1$~TeV$^{-1}$) and SM backgrounds after applying the ATLAS selection criteria, with a full emulation of the detector resolutions for the energies and angles of the outgoing photons, as well as the $\pT$-dependent reconstruction efficiencies. All distributions are generated with \gammaUPC+\mgshort, except the contribution from central exclusive production (CEP, from gluon-gluon fusion in a color-singlet exchange, $gg\to\gaga$) that is obtained with \superchic~v.3.0~\cite{Harland-Lang:2018iur}. 
A full statistical analysis of this sort of signal and background distributions for varying $m_\mathrm{G}$ values and taking into account the experimental diphoton counts observed in each mass bin, would be the basis for the alternative limits-setting method (i) described above~\cite{Schoeffel:2020svx,Baldenegro:2019whq,Baldenegro:2018hng}.

In proton-proton UPCs at the LHC, ALP constraints have been obtained in the masses range 150~GeV to 2~TeV requiring two exclusive photons produced with $m_{\gaga} \gtrsim 150$~GeV over $\lvert \eta \rvert \lesssim 2.4$ and low acoplanarity $A_{\phi}^{\gamma\gamma}< 0.01$. Given the very large backgrounds from other multiple pp pileup events, it is impossible to apply rapidity gap requirements as in the PbPb case, and the experiments require instead kinematic coincidences between the central diphoton system and one (single tagging) or both (double tagging) forward/backward protons detected in the RPs. As the forward detectors cannot get arbitrarily close to the proton beam, and the position of the LHC beam collimators limits their acceptance, the resulting coverage of the longitudinal fractional momentum loss of the protons, $\xi$, is limited. Such a requirement and all the others are summarized in Table~\ref{tab:observable-value-combined}.

For the Belle~II limits at low graviton masses, we apply the same analysis criteria used for searches for ALPs in the three-photon final state over the mass range 0.2--9.7~GeV~\cite{Belle-II:2020jti}. At least three photon candidates are considered with energy $E_{\gamma}$ above 0.65~GeV (for $m_{a}> 4$~GeV) or 1.0~GeV (for $m_{a} \le 4 $~GeV) 
and the invariant mass $m_{\gamma \gamma \gamma}$ of the three-photon is 
required to be in the range: $ 0.88\sqrt{s} \le m_{\gamma \gamma \gamma} \le 1.03\sqrt{s}$. 
As mentioned above, for BES-III, the production process is a bit different and Eq.~(\ref{eq:limitGaFromJpsi}), instead of Eq.~(\ref{eq:recast}), is employed.

%%%%%%%%%%%%%%%%%%%%%
\section{Results and discussion}
\label{sec:results}
%%%%%%%%%%%%%%%%%%%%%

\begin{figure}[!htbp]
\centering
 \includegraphics[width=0.85\textwidth]{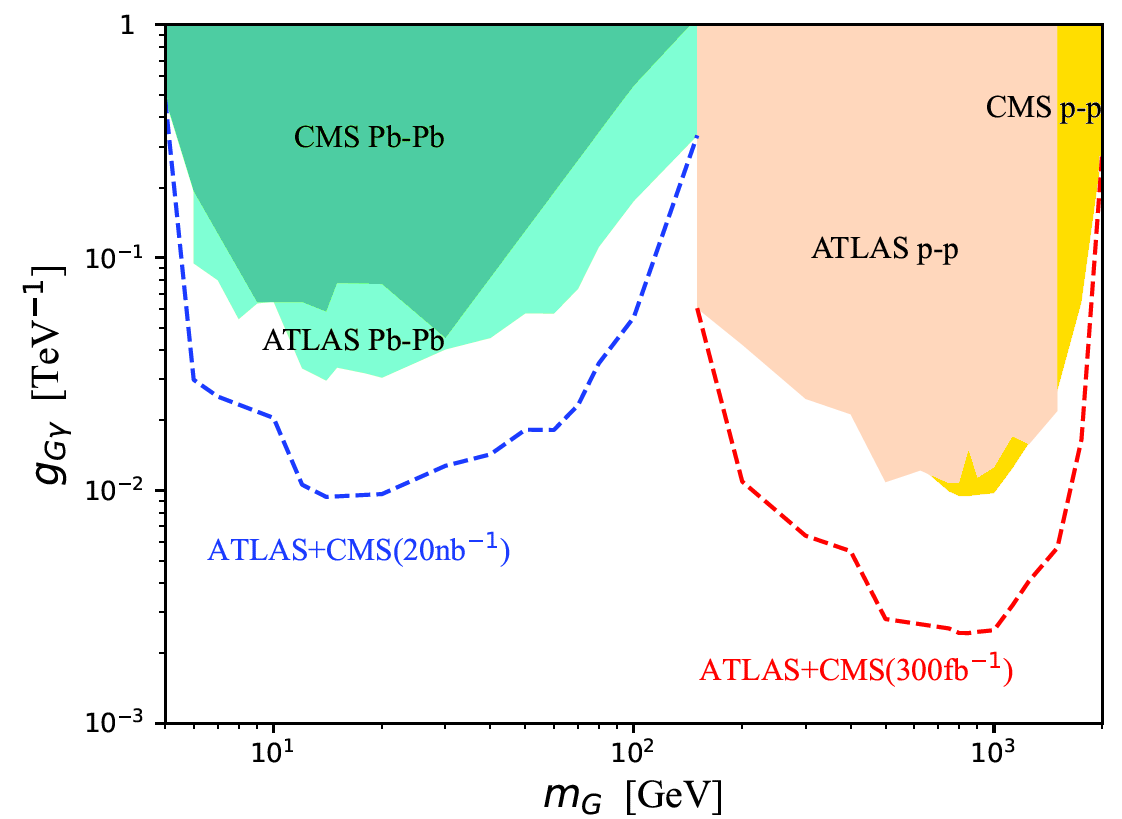}
 \caption{Exclusion limits at 95\% CL on the graviton-photon coupling as a function of the graviton mass derived from the latest ATLAS and CMS measurements of exclusive $\gaga$ production in PbPb and pp UPCs~\cite{ATLAS:2019azn,CMS:2018erd,TOTEM:2021zxa,CMS:2022zfd,ATLAS:2023zfc}.
 A photophilic scenario with $\mathcal{B}_{\mathrm{G} \to \gamma\gamma}=1$ is assumed. Extrapolated limits (dashed lines) are also shown for expected HL-LHC integrated luminosities.}
 \label{fig:br1}
\end{figure}

Using Eq.~(\ref{eq:recast}) for LHC and Belle-II, and Eq.~(\ref{eq:limitGaFromJpsi}) for BES-III, we are able to reinterpret the existing limits on the ALP-$\gamma$ coupling versus ALP mass~\cite{ATLAS:2019azn,CMS:2018erd,TOTEM:2021zxa,CMS:2022zfd,ATLAS:2023zfc,Belle-II:2020jti,BESIII:2022rzz,Baillargeon:1995dg} into the corresponding limits for graviton-$\gamma$ couplings. For the graviton limits from PbPb or pp UPCs, one can in principle keep the simplifying assumption of unity diphoton-decay branching fractions, $\mathcal{B}_{\mathrm{G},a \to \gamma \gamma} = 1$, without unitarity problems in the cross section calculations. The corresponding exclusion limits (upper limits) at 95\% confidence level (CL) for the graviton-photon coupling $g_{\mathrm{G}\gamma}=k_\gamma/\Lambda$ as a function of the mass of the graviton are displayed in Fig.~\ref{fig:br1}. %The mass range is different for lead-lead and proton-proton collisions as explained in the previous section. 
A comment is in order concerning the hypothesis $\mathcal{B}_{\mathrm{G} \to \gamma \gamma} = 1$ that, as we shall see below, is not always possible to keep. For $\mathcal{B}_{\mathrm{G} \to \gamma \gamma} < 1$, obviously, the sensitivity of the search for gravitons with exclusive diphotons decreases, due to a lower signal rate. On the other hand, the \textit{total} decay width automatically increases for decreasing $\mathcal{B}_{\mathrm{G} \to \gamma \gamma}$, but the efficiency of the search is independent of the width as it consists essentially of counting event numbers. Similarly, the region from LHC diphoton bump searches shrinks for reducing $\mathcal{B}_{\mathrm{G} \to \gamma \gamma}$ values~\cite{Jaeckel:2012yz}. Thus, there is an interplay that makes the exclusive diphotons search to gain competitiveness in the case of a broad resonance.

%%%%%%%%%%%%%%%%%%%%%%%
%%%%%%%%%%%%%%%%%%%%%%%

Let us note that the fact that no event is observed in the data at an invariant mass $m_{\gaga} = 45$~GeV~\cite{ATLAS:2023zfc} with the simulation results of Fig.~\ref{fig:massplot}, implies a direct statistical derivation (method (i) mentioned above, see Appendix~\hyperref[sec:App]{A}) of $g_{\mathrm{G} \gamma} < 4.5 \cdot 10^{-2}$~TeV$^{-1}$ at 95\% CL for $m_\mathrm{G}=45$~GeV, which is coherent with the value obtained in Fig.~\ref{fig:br1}. In Fig.~\ref{fig:br1}, we also show the limits (dashed curves) obtained by extrapolating the current results to the integrated luminosities to be recorded in PbPb and pp collisions at the HL-LHC. We take $\mathcal{L}= 20$~nb$^{-1}$ for PbPb~\cite{dEnterria:2022sut}, and a conservative $\mathcal{L}= 300$~fb$^{-1}$ for the pp case, instead of the nominal value of $\mathcal{L}= 3000$~fb$^{-1}$, given that the availability of RPs at ATLAS/CMS is not yet guaranteed over the full HL-LHC phase.

\begin{figure}[!htbp]
\centering
 \includegraphics[width=0.85\textwidth]{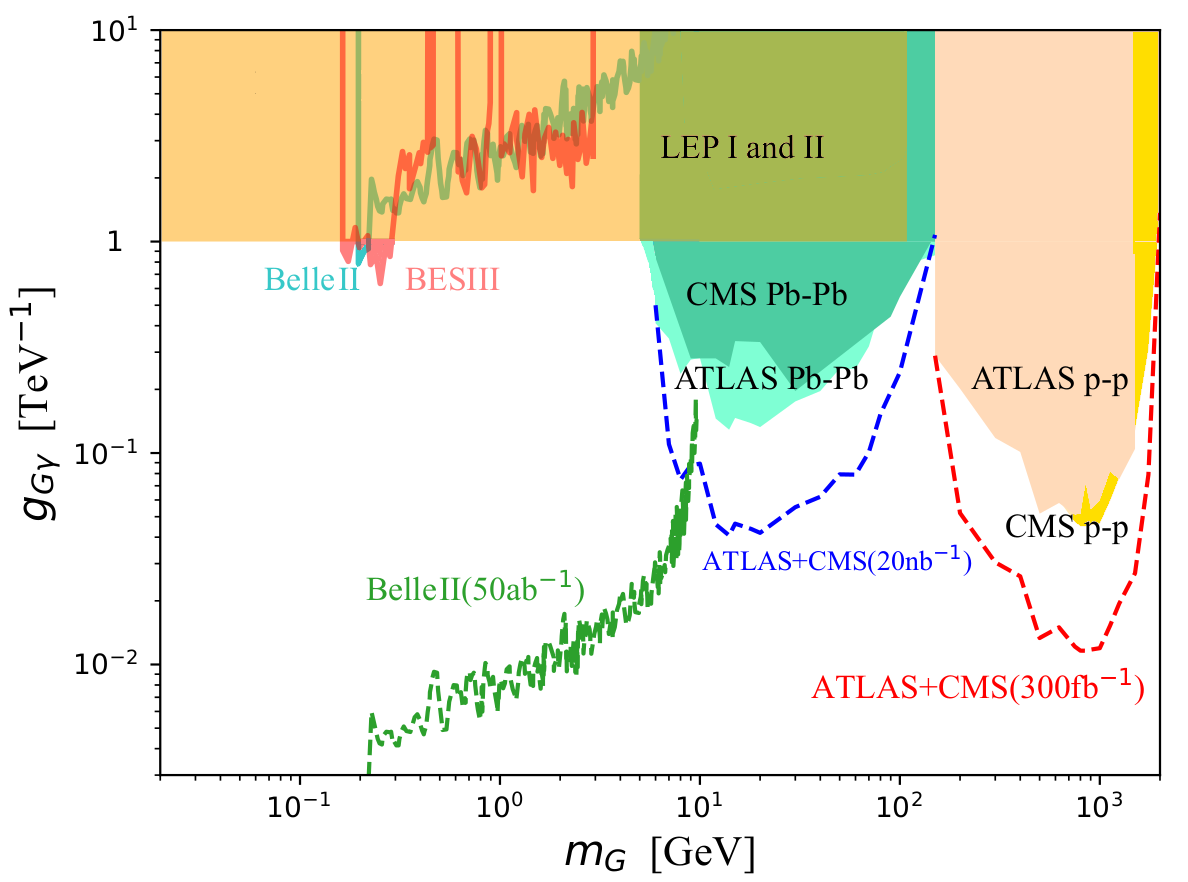}
 \caption{Exclusion limits at 95\% CL on the graviton-photon coupling as a function of the graviton mass derived from the latest ATLAS, CMS, Belle~II, BES-III, and LEP exclusive diphoton and triphoton results~\cite{ATLAS:2019azn,CMS:2018erd,TOTEM:2021zxa,CMS:2022zfd,ATLAS:2023zfc,BESIII:2022rzz,BESIII:2022rzz,Baillargeon:1995dg}. 
 A universal coupling of the graviton to SM particles is assumed, which fixes its $\gaga$ decay branching fractions as shown in Fig.~\ref{fig:diagbr} and Table~\ref{tab:branching_ratios}. Extrapolated limits (dashed lines) are also presented for the final integrated luminosities expected at Belle~II and LHC.}
 \label{fig:uc}
\end{figure}

The derivation of $g_{\mathrm{G}\gamma}$ from the measured $g_{a\gamma}$ limits at $\epem$ colliders, %Belle~II, BES-III and LEP, 
requires to consider the universal coupling scenario (Section~\ref{sec:th}), for which the diphoton branching ratio of the graviton is fixed at any given $m_\mathrm{G}$ to the values shown in Fig.~\ref{fig:diagbr} and Table~\ref{tab:branching_ratios}. Within the more realistic universal coupling approach, it becomes possible to compute the cross section of the graviton production processes at ATLAS, CMS, and $\epem$ colliders and thus to recast all ALPs limits into graviton limits using Eq.~(\ref{eq:recast}). Results are presented in Fig.~\ref{fig:uc}. Using all the experimental data, upper limits on the graviton-photon coupling are set up over $g_{\mathrm{G}\gamma}\approx 1$--0.05~TeV$^{-1}$ for masses $m_\mathrm{G} \approx 100$~MeV--2~TeV. 
Figure~\ref{fig:uc} also shows extrapolated limits (dashed curves) for the total integrated luminosities expected to be collected over the entire lifetime of the HL-LHC and Belle~II~\cite{Belle-II:2018jsg} experiments, which show that the current bounds can be improved by factors of about 100 in the low-mass region, and of 4 at high masses.

%%%%%%%%%%%%%%%%%%%%%%%
%%%%%%%%%%%%%%%%%%%%%%%

It is worth noting that the universal-coupling graviton has also a branching fraction of $\mathcal{B}_{\mathrm{G}\to \ell^+\ell^-} \approx 2.5\%$ into each pair of charged leptons, and of $\mathcal{B}_{\mathrm{G}\to \mathrm{W}^+\mathrm{W}^-} \approx 10\%$ into W$^\pm$ pairs at high masses (Table~\ref{tab:branching_ratios}). Exclusive measurements of $\gaga\to\epem,\mumu,\tautau$ in PbPb UPCs over $m_{\ell^+\ell^-}\approx 5$--100~GeV~\cite{CMS:2018erd,ATLAS:2020epq,CMS:2020skx,ATLAS:2020mve,ATLAS:2022ryk,CMS:2022arf} and in pp UPCs over $m_{\ell^+\ell^-}\approx 100$--1000~GeV~\cite{CMS:2011vma,ATLAS:2015wnx,ATLAS:2017sfe,CMS:2018uvs}, as well as of $\gaga\to\mathrm{W}^+\mathrm{W}^-$ in pp UPCs over $m_{\mathrm{W}^+\mathrm{W}^-}\approx 160$--2000~GeV~\cite{CMS:2013hdf,CMS:2016rtz,ATLAS:2016lse,ATLAS:2020iwi,CMS:2022dmc}, have not shown any significant excess with respect to the SM predictions. Such measurements are in agreement with (but less stringent than) the graviton limits derived from the exclusive diphoton measurements discussed here. 

A discussion is also in place regarding the comparison of our massive graviton limits with those set by other inclusive searches at the LHC. 
%Massive spin-2 particles are typically predicted by BSM models proposed to explain the very large gap between the electroweak ($10^2$~GeV) and Planck ($10^{19}$~GeV) scales (``hierarchy problem'') based on the existence on new compact spatial dimensions. Graviton-like particles appear as Kaluza--Klein (KK) excitations of these extra dimensions %\footnote{The model differences arise from the number of extra dimensions considered, and their compactification.} in the Randall--Sundrum (RS)~\cite{Randall:1999ee}, and  Arkani-Hamed--Dimopoulos--Dvali (ADD)~\cite{Arkani-Hamed:1998sfv} approaches (with model differences arising mostly from the number of extra dimensions considered, and their compactification). 
As mentioned in the introduction, 
Graviton-like particles appear as Kaluza--Klein  excitations of  extra dimensions in the RS~\cite{Randall:1999ee}, and  ADD~\cite{Arkani-Hamed:1998sfv} approaches (with model differences arising mostly from the number of extra dimensions considered, and their compactification). 
Both RS and ADD gravitons have been searched for in standard parton-parton collisions at the LHC, in the form of high-mass dijet, dilepton, and/or diphoton resonances, $\mathrm{pp}\to \mathrm{G} \to j j,\ell^+\ell^-,\gaga$, above the corresponding dominant perturbative quantum chromodynamics (pQCD) continuum backgrounds, $\mathrm{pp}\to j j,\,\ell^+\ell^-,\,\gaga+X$. 
In the high mass range, our universal- coupling scenario predicts gravitons predominantly decaying into two high-$\pT$ hadronic jets with $\mathcal{B}_{\mathrm{G}\to jj} \approx 65\%$ (Fig.~\ref{fig:diagbr}). 
To date, exploiting the full Run-2 integrated luminosities (140~fb$^{-1}$) of pp collisions at 13~TeV, no localized dijet excess has been found up to a few TeV~\cite{CMS:2019gwf,ATLAS:2020iwa,CMS:2022usq}. 
For a graviton mass of 1~TeV, our limits predict $g_{\mathrm{G} \gamma} \lesssim 4.5 \cdot 10^{-2}$~TeV$^{-1}$ which, using the branching fraction of the graviton decaying into two jets, would translate into a production cross section smaller than 0.44~pb for the $\mathrm{pp}\to \mathrm{G} \to j j$ process at 13~TeV~\cite{Das:2016pbk}. However, the pQCD cross section for $\mathrm{pp} \to j j $ at 13~TeV at $m_{jj}=1$~TeV (with the difference in rapidities of the two jets being smaller than 1.2) has been measured to be more than 200 times larger, $\mathcal{O}(100)$~pb within a few \% total uncertainty~\cite{ATLAS:2018qto}. At lower masses, the situation is even more dire, with pQCD dijet background invariant cross sections per mass bin increasing as a power law with exponent $n\approx 5$. This explains why such a potential graviton would not have been observed in inclusive dijet searches, and that the exclusive photon-fusion-based search presented in this paper is competitive in the broader range of masses covered.

\begin{figure}[!htbp]
\centering
 \includegraphics[width=0.75\textwidth]{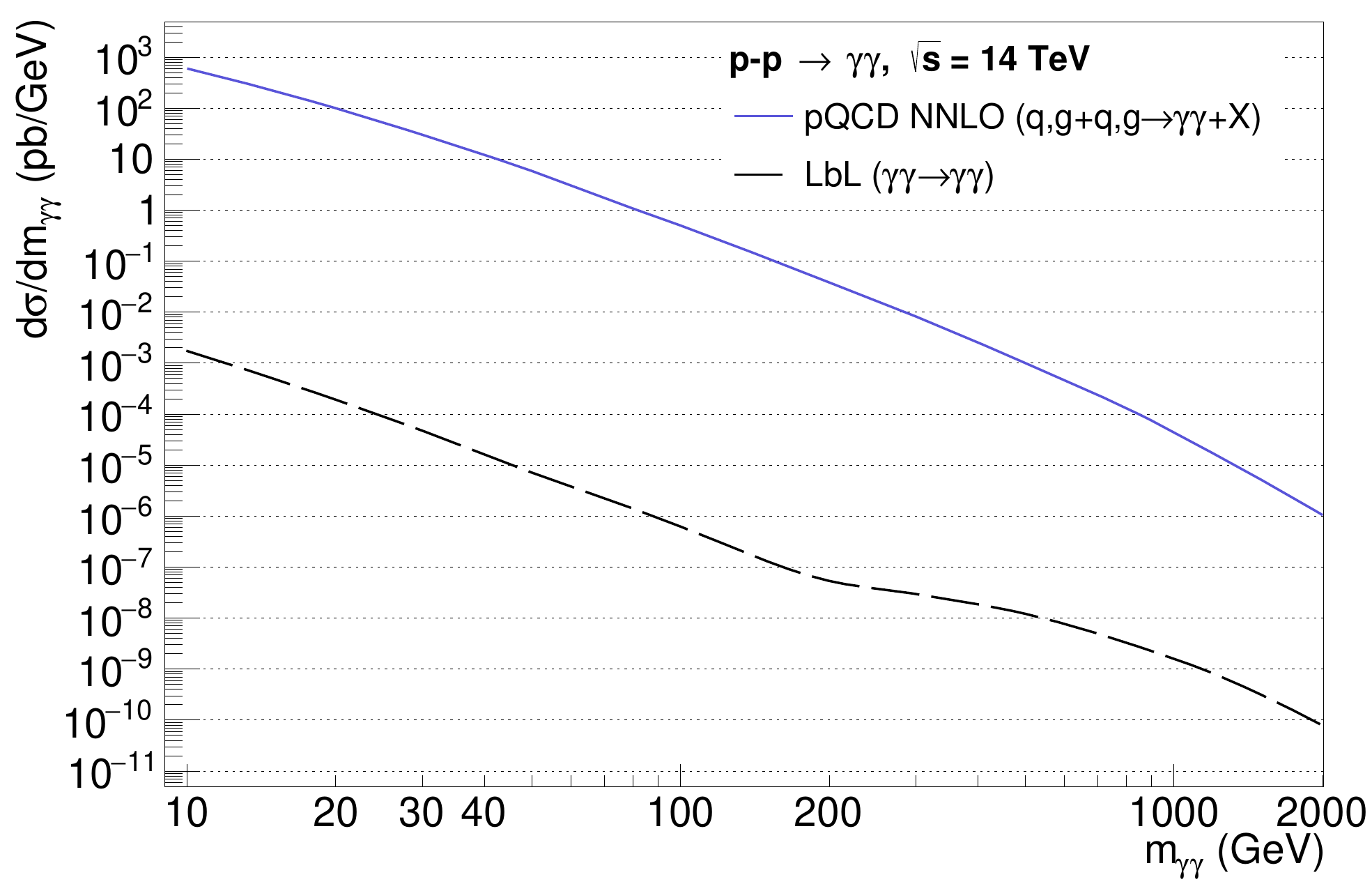}
 \caption{Cross sections for continuum NNLO ($\mathrm{pp}\to\gaga+X$) and exclusive LbL ($\mathrm{pp}\overset{\gamma\gamma}{\to}\mathrm{p}\gaga\mathrm{p}$) diphoton backgrounds as a function of mass, for proton-proton collisions at $\sqrts = 14$~TeV.}
 \label{fig:sigma_aa}
\end{figure}

Similar searches for the inclusive production of RS and ADD gravitons have been performed in the diphoton channel in pp collisions at the LHC, $\mathrm{pp}\to \mathrm{G} \to \gaga$. No diphoton spin-2 resonance excess has been neither found above the inclusive pQCD diphoton background, and exclusion limits for RS gravitons have been set by both ATLAS and CMS over $m_\mathrm{G} \approx 100$--3000~GeV~\cite{CMS:2018dqv,ATLAS:2021uiz}. Since inclusive searches for diphoton resonances include, by definition, also any potential $\gaga\to \mathrm{G}\to \gaga$ production, the reader may wonder what advantage the exclusive searches presented here provide in terms of limits settings. First, the exclusive final states in UPCs can probe much lower diphoton masses without pileup and collision backgrounds that prevent photon isolation in inclusive searches. Second, any exclusive $\gaga$ graviton searches are complementary to the inclusive ones, as they have different sources of systematic (experimental and theoretical) uncertainties. Third, arguably the clearer advantage is in the very different sizes of the irreducible backgrounds as shown in Fig.~\ref{fig:sigma_aa}, which compares the cross sections for the continuum pQCD ($\mathrm{pp}\to\gaga+X$) and exclusive LbL ($\mathrm{pp}\overset{\gamma\gamma}{\to}\mathrm{p}\gaga\mathrm{p}$) diphoton backgrounds as a function of mass for proton-proton collisions at $\sqrts = 14$~TeV. The parton-induced pQCD curve has been obtained at LO with \mgshort\ and scaled up by a $K$-factor of $K\approx 4$--2, at low and high masses respectively, derived from next-to-next-to-leading-order (NNLO) calculations~\cite{Catani:2018krb,Gehrmann:2020oec}. The LbL curve has been computed with \gammaUPC+\mgshort\ as explained in Section~\ref{sec:th}, the local ``bump'' at $m_{\gaga}\approx 350$~GeV is due to the onset of top-antitop quark boxes (aka.\ the resonant anomalous threshold~\cite{Passarino:2018wix}). This figure shows that the cross sections for inclusive $\gaga$ are up to 6 orders-of-magnitude larger than the exclusive $\gaga$ ones: At $m_{\gaga}\approx 1$~TeV, $\mathrm{d}\sigma(\mathrm{pQCD,LbL})/\mathrm{d}m_{\gaga}\approx 50$~ab/GeV,~1~zb/GeV, respectively. Namely, the exclusive graviton $\gaga$ production and decay mode considered in this paper is subject to negligible SM irreducible backgrounds, and with proper control of instrumental effects (and for equal integrated luminosities) the G-$\gamma$ coupling limits that can be set from exclusive analyses can be more competitive than those from standard inclusive graviton searches at the LHC. This is particularly true for potentially nonresonant gravitons (or with a width much larger than the detector diphoton resolution), where the signal would be further washed out and swamped by the pQCD background in inclusive searches, but would still appear as an excess over the negligible LbL cross section in exclusive studies. 

%%%%%%%%%%%%%%%%%%%%%%%
%%%%%%%%%%%%%%%%%%%%%%%

%%%%%%%%%%%%%%%%%%%%%
\section{Conclusions}
\label{sec:conclusion}
%%%%%%%%%%%%%%%%%%%%%

We have examined the possibility of searching for massive spin-2 (graviton) particles produced via two-photon processes and decaying back to photons ($\gaga\to\mathrm{G}\to\gaga$), in ultraperipheral collisions (UPCs) of lead ions, $\mathrm{PbPb}\overset{\gamma\gamma}{\to}\mathrm{Pb}\,\mathrm{G(\gaga)}\,\mathrm{Pb}$, and of protons, $\mathrm{pp}\overset{\gamma\gamma}{\to}\mathrm{p}\,\mathrm{G(\gaga)}\,\mathrm{p}$, at the LHC, as well in three-photon final states in $\epem$ collisions measured at the Belle~II, BES-III, and LEP experiments, $\epem \to \mathrm{G}(\gaga)\gamma$. We have considered a minimal effective field theory model that describes a linearized kinetic Lagrangian for a spin-2 graviton and its coupling to all standard model particles. Such a universal-coupling graviton model allows to consider a free G-$\gamma$ coupling for the case of $\epem$ collisions with three-photon final states without breaking the perturbative unitarity of the calculations.  
Based on similar searches performed for pseudoscalar axion-like particles (ALPs), and taking into account the different cross sections, $\gaga$ partial widths, and decay kinematics of the pseudoscalar and tensor particles, we can reinterpret existing experimental bounds on the ALP-$\gamma$ coupling into G-$\gamma$ ones. 
With this goal, simulations have been run for graviton and ALPs samples, reproducing the experimental searches for diphoton and triphoton excesses. 
For PbPb and pp collisions, 95\% CL upper limits $g_{\mathrm{G}\gamma}\approx 1$--0.1~TeV$^{-1}$ have been set over $m_\mathrm{G} = 5$~GeV to 100~GeV, and over $g_{\mathrm{G}\gamma}\approx 0.5$--0.05~TeV$^{-1}$ for $m_\mathrm{G} = 150$~GeV and 2~TeV, respectively. Compared to standard inclusive searches of high-mass diphoton bumps above the pQCD continuum at the LHC, the exclusive UPC final states benefit from reduced pileup backgrounds, negligible SM model irreducible continuum backgrounds, and the possibility to probe graviton masses in the few-GeV range. The $\epem$ measurements allow further constraining the graviton-photon coupling down to $g_{\mathrm{G} \gamma} \approx 1$~TeV$^{-1}$ at even smaller graviton masses, from 100~MeV up to about 10~GeV.
Such bounds can be improved by factors of 100 at Belle~II in the low-mass region, and of 4 at the HL-LHC at high masses, with their expected full integrated luminosities.\\

\paragraph*{Acknowledgments---} %Discussions with ... are gratefully acknowledged. 
Support from the European Union's Horizon 2020 research and innovation program (grant agreement No.824093, STRONG-2020, EU Virtual Access ``NLOAccess''), the ERC grant (grant agreement ID 101041109, ``BOSON"), and the French ANR (grant ANR-20-CE31-0015, ``PrecisOnium''), are acknowledged.

%\vfill
\clearpage
\appendix
\section*{Appendix A: Statistical equivalence of limit-setting procedures (i) and (ii) of Section~\ref{sec:analysis}.}
\label{sec:App}

In order to derive an exclusion limit on the signal cross section and its associated coupling (with $\sigma_g = g^2 \sigma_{g \equiv 1}$), we need to assume a set of observed data. As commonly done, we assume that no statistical fluctuations are present in these pseudodata, which are usually dubbed ``Asimov'' data and that we denote here with a prime. As experimental data are used to derive the limits, the collected integrated luminosity $\mathcal{L}_d$ is considered in this discussion.

The observed events follow a Poisson distribution, and to simplify the discussion we can neglect the systematic uncertainties here. The statistical size of the event data sample, together with the prediction for the event rates, define the likelihood function needed:
\be
{ L}(\sigma )=\mathrm{Pr}( n' |b+\sigma \mathcal{L}_d) \mbox{ with } \mathrm{Pr}( \hat n |n)=\frac{ {n^{\hat n}} e^{-n } }{\hat n !}. 
\label{eq:L}
\ee
Here $\mathrm{Pr}( \hat n |n)$ is the probability density function of finding 
$\hat n$ events if $n$ ($n'$) events are expected in the domain selected after experimental requirements, and $b$ is the expected number of events from the background, given here mostly by the SM LbL prediction. We aim to obtain projected exclusion limits at $95\%$ CL. Then, we define the posterior probability density for $\sigma$ as $ { L}(\sigma ) \pi(\sigma)$ where the prior is $\pi(\sigma)=1$ if $\sigma>0$, and $0$ otherwise. In order to derive the limits, we assume that no event is observed, \textit{i.e.} $n'=0$, with the consequence that an upper bound on the signal event rate can be set. The higher posterior density region at $1-\alpha $ 
credibility level is solved analytically and is simply given by:
\be
1-\alpha=\frac{\int^{\sigma_\alpha}_0 L(\sigma) \pi(\sigma)}
{\int_{0}^\infty L(\sigma)\pi(\sigma)}= 1-e^{-\sigma_\alpha \mathcal{L}_d}.
\ee
This gives the upper limit cross section for the signal: 
\be
\sigma_\alpha=-\frac{1}{\mathcal{L}_d} \log(\alpha).
\ee
Then for a $95\%$ credible interval, we take $\alpha=0.05$ and the exclusion limit is simply given by $\sigma_\alpha \approx 3 \mathcal{L}_d^{-1}$.
This implies that the corresponding upper limit on the ALP-photon coupling $g_{a\gamma}$ is given by:
\be
g_{a\gamma} = \sqrt{\frac{\sigma_\alpha}{\sigma_{a,\text{gen}}}} \ g_{a\gamma, \text{gen}}.
\label{eq:toto1}
\ee
Here, $\sigma_{a,\text{gen}}$ is the generated cross section for the ALPs production and $g_{a\gamma, \text{gen}}$ the corresponding ALP-$\gamma$ coupling. Obviously, we can perform the same exercise for the graviton-$\gamma$ coupling, leading to:
\be
g_{\mathrm{G} \gamma} = \sqrt{\frac{\sigma_\alpha}{\sigma_{\mathrm{G},\text{gen}}}} \ g_{\mathrm{G} \gamma, \text{gen}}.
\label{eq:toto2b}
\ee
Then, the ratio of Eqs.~(\ref{eq:toto1}) and (\ref{eq:toto2b}) gives Eq.~(\ref{eq:recast}) of the method (ii) discussed in Section~\ref{sec:analysis}, as expected.

%%%%%%%%%%%%%%%
%\section*{Acknowledgements}

%%%%%%%%%%%%%%%

%%%%%%%%%%%%%%%%%%%%%
\clearpage
\bibliographystyle{myutphys}
\bibliography{biblio.bib}

\providecommand{\href}[2]{#2}\begingroup\raggedright

\end{document}